\documentclass[twocolumn,english,aps,pra]{revtex4}

\usepackage[latin1]{inputenc}
\usepackage{graphicx}
\usepackage{amssymb}

\makeatletter

\usepackage{graphicx}

\usepackage{graphicx}

\usepackage{babel}
\makeatother

\begin{document}

%
%
%

\title{Semiclassical Wigner distribution for two-mode 
       entangled state \\ generated by an optical parametric 
       oscillator}

\author{K. Dechoum} 
\email{kaled@if.uff.br}

\author{M. D. Hahn} 

\author{A. Z. Khoury}
\email{khoury@if.uff.br}
\affiliation{  Instituto de F\'{\i}sica, 
               Universidade Federal Fluminense,\\ 
               Av. Gal. Milton Tavares de Souza s/n, 
               24210-346 Niter\'oi - RJ, Brazil}

\author{R. O. Vallejos}
\email{vallejos@cbpf.br}
\affiliation{ Centro Brasileiro de Pesquisas F\'{\i}sicas,  
              Rua Dr.~Xavier Sigaud 150, \\
              22290-180 Rio de Janeiro - RJ, Brazil}

\date{\today}

\begin{abstract}

We derive the steady state solution of the Fokker-Planck equation that 
describes the dynamics of the nondegenerate optical parametric oscillator 
in the truncated Wigner representation of the density operator. 
The adiabatic limit of strong pump damping is assumed. 
This phase space image provides a clear view of the intracavity two-mode 
entangled state valid in all operating regimes of the OPO. 
A nongaussian distribution is obtained for the above threshold solution. 
\end{abstract} \pacs{} \maketitle

\section{Introduction}

Among the quasiprobability functions that represent the density operator 
of a quantum state, the Wigner distribution has undoubtedly advantages over 
others, since in this phase space representation the quantum-classical 
correspondence is, in general, much more visible. At the same time these 
functions contain all information available in the density operator.  

Unfortunately, just a few states have been represented by this distribution 
due to the difficulty to solve Fokker-Planck equations for nonlinear systems. 
Indeed, the phase space description of quantum systems is well known for 
quadratic hamiltonians, but very little is known for nonlinear 
systems, such as, for example, the single-mode degenerate parametric oscillator 
\cite{kinsler}, or the transverse multimode degenerate parametric oscillator 
\cite{kaled2}. 
 
Of special interest is the two-mode entangled state generated in the optical 
parametric oscillator (OPO), used in many experiments of quantum information. 
The Wigner distribution of this state was obtained exactly using the solution 
derived from the complex P-representation \cite{karen}, and tranformed into a 
Wigner distribution. 
Since this analytical result appears as an infinite series of gamma functions 
(hypergeometric series), some important physical aspects are hidden by its 
mathematical complexity. For example, two-mode entanglement is not of easy 
identification from the usual partial transpose criterion \cite{simon,duan}.
The same happens to the generalized P-representation for the intracavity 
parametric oscillator derived in Ref.\cite{mcneil}. In order to calculate 
the intensity correlations, the authors obtained an integral expression 
that depends on degenerate hypergeometric functions, and the outcome also 
depends on an infinite sum. 

In the present work, we derive the Wigner function of a two-mode quantum 
state of the electromagnetic field generated by an OPO using a truncated 
Wigner equation. This truncated approximation is valid when the nonlinearity 
is small enough, which is true for most of the OPO experiments developed 
so far. A simple expression is obtained for the Wigner distribution, which 
renders clear the many quantum features of the intracavity OPO field 
such as squeezing and entanglement.
Moreover, this expression is valid in all regions of the OPO operation regimes, 
that is, below, and above threshold, where the validity of both the linear 
approximation, and the perturbation theory break down. This distribution also 
describes the OPO quantum fluctuations properly around the operation threshold, 
while the usual linear approach provides divergent results. 
In order to illustrate the reliability of the Wigner function we obtain,
moments calculated from this distribution are compared with the
exact quantum results, for any values of the control parameters of the
OPO.

\section{Theoretical model}

We present here a model of three quantized modes coupled by a nonlinear crystal 
inside a triply resonant Fabry-Perot cavity.
The Heisenberg picture Hamiltonian that describes this open system is
given by\cite{carmichael}
\begin{eqnarray}
\hat{H} &=& \sum_{i=0}^{2} \hbar \omega_{i} \hat{a}_{i}^{\dagger} \hat{a}_{i}
+ i \hbar \chi \left(\hat{a}_{1}^{\dagger} \hat{a}_{2}^{\dagger}
\hat{a}_{0} - \hat{a}_{1} \hat{a}_{2} \hat{a}_{0}^{\dagger} \right) \nonumber \\
&+&  i\hbar \left( E e^{-i\omega_{0} t}\hat{a}_{0}^{\dagger} -
E^{*} e^{i\omega_{0} t}\hat{a}_{0} \right) \nonumber
\\
&+&\sum_{i=0}^{2} \left(
\hat{a}_{i} \hat{\Gamma}_{i}^{\dagger} + \hat{a}_{i}^{\dagger}
\hat{\Gamma}_{i} \right)\;.
\label{1}
\end{eqnarray}

Here $E$ represents the external coherent driving pump field at frequency 
$\omega_{0}$. The operators $\hat{a}_{0}$, $\hat{a}_{1}$ and $\hat{a}_{2}$ 
represent the pump, signal and idler fields, respectively, satisfying the 
following frequency matching condition, $\omega_{0}=\omega_{1}+\omega_{2}$. 
The terms $\hat{\Gamma}_{i}$ represent damping reservoir operators, 
and $\chi$ is the nonlinear coupling constant due to the second order 
polarizability of the nonlinear crystal.

The master equation for the reduced density operator, after the elimination 
of the heat bath by standard techniques \cite{carmichael}, is given by
\begin{eqnarray}
\frac{\partial \hat{\rho}}{\partial t} &=& 
-i \sum_{i=0}^{2} \omega_{i} \left[ \hat{a}_{i}^{\dagger}\hat{a}_{i}, \hat{\rho} \right] +
 \chi\left[ \hat{a}_{1}^{\dagger} \hat{a}_{2}^{\dagger}\hat{a}_{0}, 
\hat{\rho} \right] - 
 \chi\left[ \hat{a}_{1} \hat{a}_{2} \hat{a}_{0}^{\dagger},\hat{\rho} \right] \nonumber \\
&&+ E e^{-i\omega_{0} t}\left[ \hat{a}_{0}^{\dagger},\hat{\rho} \right]- 
 E^{*} e^{i\omega_{0} t}\left[ \hat{a}_{0},\hat{\rho} \right]\nonumber \\
&&+\sum_{i=0}^{2} \gamma_{i} \left( 2 \hat{a}_{i} \hat{\rho} 
\hat{a}_{i}^{\dagger} - \hat{a}_{i}^{\dagger} \hat{a}_{i} \hat{\rho} -
\hat{\rho} \hat{a}_{i}^{\dagger} \hat{a}_{i} \right)\;,
\label{2}
\end{eqnarray}

\noindent where $\gamma_{i}$ is the corresponding mode damping rate. 

In order to treat the operators evolution, we now turn to the method of 
operator representation theory. These techniques can be used to transform
the density matrix equation of motion into c-number Fokker-Planck or 
stochastic equations.
We can write down the master equation in the Wigner representation, by using 
the following characteristic function
\begin{eqnarray}
\chi_{W} (z,z^{*}) &=& Tr \left( \rho e^{i z^{*} a^{\dagger} + i z a} \right) 
\nonumber\\
&=& Tr \left( \rho e^{i z^{*} a^{\dagger}} e^{i z a} e^{-|z|^{2}/2} \right)\;,
\label{3}
\end{eqnarray}
\noindent so that the Wigner distribution can be written as Fourier transform 
of the characteristic function:
\begin{equation}
W(\alpha ,\alpha^{*}) = \frac{1}{\pi^{2}} \int_{-\infty}^{\infty} d^{2} z \; 
\chi_{W}(z,z^{*}) e^{-iz^{*}\alpha^{*}} e^{-iz\alpha}\;.
\label{4}
\end{equation}

The phase space Wigner equation for the nondegenerate parametric amplifier 
that corresponds to the master equation (\ref{2}) is then
\begin{eqnarray}
\frac{\partial W}{\partial t}&=& 
\left\{ 
\frac{\partial}{\partial \alpha_{0}} \left( i \omega_{0} \alpha_{0}+ 
\gamma_{0} \alpha_{0} + \chi \alpha_{1} \alpha_{2} - E e^{-i\omega_{0}t} 
\right)\right. \nonumber \\
&+& \frac{\partial}{\partial \alpha_{0}^{*}} 
\left(-i \omega_{0} \alpha_{0}^{*} + \gamma_{0} \alpha_{0}^{*}+
\chi \alpha_{1}^{*} \alpha_{2}^{*} - E^{*} e^{i\omega_{0}t}
\right)  \nonumber \\  
&+& \frac{\partial}{\partial \alpha_{1}} 
\left( i \omega_{1} \alpha_{1}+
\gamma_{1} \alpha_{1} - \chi \alpha_{2}^{*} \alpha_{0} \right) \nonumber \\
&+& \frac{\partial}{\partial \alpha_{1}^{*}} 
\left(-i \omega_{1} \alpha_{1}^{*} + 
\gamma_{1} \alpha_{1}^{*} - \chi \alpha_{2} 
\alpha_{0}^{*} \right)  \nonumber \\
&+& \frac{\partial}{\partial \alpha_{2}} 
\left( i \omega_{2} \alpha_{2}+
\gamma_{2} \alpha_{2} - \chi \alpha_{1}^{*} \alpha_{0} \right) \nonumber \\
&+& \frac{\partial}{\partial \alpha_{2}^{*}} 
\left(-i \omega_{2} \alpha_{2}^{*} +  
\gamma_{2} \alpha_{2}^{*} - \chi \alpha_{1} 
\alpha_{0}^{*} \right)  \nonumber \\
&+& \gamma_{0} \frac 
{\partial^{2}}{\partial \alpha_{0} \partial \alpha_{0}^{*}} +  
\gamma_{1} \frac 
{\partial^{2}}{\partial \alpha_{1} \partial \alpha_{1}^{*}} +  
\gamma_{2} \frac 
{\partial^{2}}{\partial \alpha_{2} \partial \alpha_{2}^{*}}  \nonumber \\ 
&+& \left. \frac{\chi}{4} \left(
\frac{\partial^{3}}{\partial \alpha_{1} \partial \alpha_{2} 
\partial \alpha_{0}^{*}} +  
\frac{\partial^{3}}{\partial \alpha_{1}^{*}\partial 
\alpha_{2}^{*} \partial \alpha_{0}} \right)
\right\} W\;.
\label{5}
\end{eqnarray}  

This is not a Fokker-Planck equation due to the third order derivative term, but 
in the case where $\chi$ is small enough we can drop this term. The truncated equation 
so obtained is a genuine Fokker-Planck equation with a positive diffusion term, 
and we can easily derive the corresponding stochastic differential equations
in the rotating frame ($\tilde{\alpha}_j = \alpha_j\exp{(-i\omega_j t)}$ with $j=0,1,2$):
\begin{eqnarray}
&& \frac{d \tilde{\alpha}_{0}}{dt} = -\gamma_{0} \tilde{\alpha}_{0} + E - 
\chi \tilde{\alpha}_{1} \tilde{\alpha}_{2} + \sqrt{\gamma_{0}} \xi_{0}(t)
\;,\nonumber \\ 
&& \frac{d \tilde{\alpha}_{1}}{dt} = -\gamma_{1} \tilde{\alpha}_{0} + 
\chi \tilde{\alpha}_{0} \tilde{\alpha}_{2}^{*} + \sqrt{\gamma_{1}} \xi_{1}(t) 
\;,\nonumber \\
&& \frac{d \tilde{\alpha}_{2}}{dt} = -\gamma_{2} \tilde{\alpha}_{0} + 
\chi \tilde{\alpha}_{0} \tilde{\alpha}_{1}^{*} + \sqrt{\gamma_{2}} \xi_{2}(t) \;.
\end{eqnarray}
It is possible to find a stationary solution of the truncated Fokker-Planck 
equation when we adiabatically eliminate the pump mode variables. This means 
that we are considering the relaxation rate of this mode much larger than 
those of the downconverted modes, that is 
$\gamma_{0} \gg \gamma_{1},\gamma_{2}$. The stationary solution for the pumped 
mode is 
\begin{equation}
\tilde{\alpha}_{0}= \frac{1}{\gamma_{0}} \left[ E - \chi \tilde{\alpha}_{1} 
\tilde{\alpha}_{2} + \sqrt{\gamma_{0}} \xi_{0} (t) \right]\;,
\label{8}
\end{equation}
where the noise term is retained in the adiabatic elimination in order to properly 
deal with the noise dynamics. We are then left with two nonlinear dynamical equations
for the complex amplitudes of the down-converted fields, where the pump amplitude 
is replaced by expression (\ref{8}).

We now define the following real quadrature variables:
\begin{eqnarray}
x_{1} =  \tilde{\alpha}_{1} + \tilde{\alpha}_{1}^{*}\;, \,\,\,\,\,\,\,\,\,\, 
y_{1} = -i\left( \tilde{\alpha}_{1} - \tilde{\alpha}_{1}^{*} \right )\;,
\nonumber \\
x_{2} =  \tilde{\alpha}_{2} + \tilde{\alpha}_{2}^{*}\;, \,\,\,\,\,\,\,\,\,\, 
y_{2} = -i\left( \tilde{\alpha}_{2} - \tilde{\alpha}_{2}^{*} \right )\;,
\end{eqnarray}
so that the remaining dynamical equations can be cast in the compact form:
\begin{equation}
\frac{d \mathbf{X}}{dt} = \mathbf{A} + \mathbf{B}\,\mathbf{\xi}(t)
\end{equation}
where $\mathbf{X}=[x_1,y_1,x_2,y_2]^T$ is a column vector, and the drift matrix is 
defined as 
\begin{equation}
\mathbf{A} = \gamma \left[
\begin{array}{c}
-x_{1} + \mu x_{2} - \frac{g^{2}}{2} x_{1} \left(x_{2}^{2} +y_{2}^{2}\right)  \\
\noalign{\medskip}
-y_{1} - \mu y_{2} - \frac{g^{2}}{2} y_{1} \left(x_{2}^{2} +y_{2}^{2}\right)  \\
\noalign{\medskip}
-x_{2} + \mu x_{1} - \frac{g^{2}}{2} x_{2} \left(x_{1}^{2} +y_{1}^{2}\right)  \\
-y_{2} - \mu y_{1} - \frac{g^{2}}{2} y_{2} \left(x_{1}^{2} +y_{1}^{2}\right)  \\
\end{array}
\right]\;.
\end{equation}
We have set $\gamma_{1}=\gamma_{2}=\gamma$, which is a reasonable physical assumption 
for most OPO experiments, and defined the normalized pump parameter $\mu = \chi E/(\gamma \gamma_{0})$, 
as well as the nonlinear coupling $g = \chi/(\sqrt{2 \gamma \gamma_{0}})$.  
$\mathbf{B}$ is a $4\times 6$ matrix defined as
\begin{equation}
\mathbf{B} = \sqrt{2 \gamma} \left(
\begin{array}{cccccc}
1 &0 &0 &0 & \frac{g}{\sqrt{2}}x_{2} & \frac{g}{\sqrt{2}}y_{2} \\
0 &1 &0 &0 & -\frac{g}{\sqrt{2}}y_{2} & \frac{g}{\sqrt{2}}x_{2} \\
0 &0 &1 &0 & \frac{g}{\sqrt{2}}x_{1} & \frac{g}{\sqrt{2}}y_{1} \\
0 &0 &0 &1 & -\frac{g}{\sqrt{2}}y_{1} & \frac{g}{\sqrt{2}}x_{1} \\
\end{array}
\right)
\end{equation}
and $ {\mathbf \xi}(t) $ is a six component column vector whose entries are uncorrelated real 
Gaussian noises associated with the noise terms for the three interacting fields.

Now this set of stochastic equations can be mapped onto a genuine Fokker-Planck equation:
\begin{equation}
\frac{\partial W(\mathbf{X})}{\partial t} = \left[ -\frac{\partial}{\partial \mathbf{X}_{i}} \mathbf{A}_{i} + 
\frac{1}{2}\frac{\partial}{\partial \mathbf{X}_{i}} \frac{\partial}{\partial \mathbf{X}_{j}} \mathbf{D}_{ij}\right]
W(\mathbf{X})
\label{10}
\end{equation}
where the diffusion matrix is defined as $\mathbf{D}= \mathbf {B} \mathbf {B}^{T}$, and summation over repeated 
indices is assumed throughout the text.

This Fokker-Planck equation admits a steady state solution in the potential form 
$W(\mathbf{X})= N \exp{(-\int \mathbf{Z}_{i}\,d\mathbf{X}_{i})}$ with $\mathbf{Z}_{i}$ given by
\begin{equation}
\mathbf{Z}_{i} = \mathbf{D}_{ik}^{-1} \left[2 \mathbf{A}_{k} - \frac{\partial}{\partial \mathbf{X}_{j}}
\mathbf{D}_{ij} \right]\;.
\end{equation}

This solution provides a quite simple form for the Wigner distribution which allows for the calculation of 
several statistical properties. It is important to notice that the potential solution is achievable only 
when $\gamma_1=\gamma_2\equiv\gamma$ as we have already assumed \cite{gardiner}. 
In this case, the steady state Wigner distribution for signal and idler reads:
\begin{eqnarray}
&& W(x_{1},y_{1},x_{2},y_{2}) = {\cal{N}} \exp \left\{\frac{-1/2}
{1+\frac{g^2}{2}\left( x_{1}^{2}+ y_{1}^{2}+x_{2}^{2}+ y_{2}^{2}\right)}
\right. 
\nonumber\\
&& \times\left[ \left( 1+g^2\right)\left(x_{1}^{2}+ y_{1}^{2}+x_{2}^{2}+ y_{2}^{2}\right) + 
2 \mu \left( y_{1}y_{2} -x_{1}x_{2} \right)\right. 
\nonumber \\
&& + \left. \left. g^{2} \left(x_{1}^{2} +y_{1}^{2}\right) 
\left(x_{2}^{2} +y_{2}^{2}\right) \right] \right\} 
\label{11}
\end{eqnarray}
where ${\cal{N}}$ is the normalization constant and unimportant terms $O(g^4)$ were 
neglected in the argument of the potential solution.

As a concrete physical example, expression (\ref{11}) can describe the 
joint Wigner distribution for two polarization modes of a frequency 
degenerate type II OPO, where signal and idler are distinguished by 
their polarization states. 
All statistical properties, including experimentally accessible quantities 
like quadrature noise and correlations in the stationary state may be 
calculated with this distribution in any OPO operation regime. 
It is interesting to notice that the distribution given by 
expression (\ref{11}) is single peaked below threshold 
operation ($\mu < 1$) and becomes double peaked above threshold. 
In Fig. (1), it is possibible to visualize the conditional Wigner 
distribution for fixed values of $y_{1}$ and $y_{2}$, for an 
OPO operating below, at, and above threshold. It is easy to identify 
the squeezed and unsqueezed quadratures in all regimes.

\begin{figure}
\resizebox*{8.0cm}{5cm}{\rotatebox{0}{\includegraphics{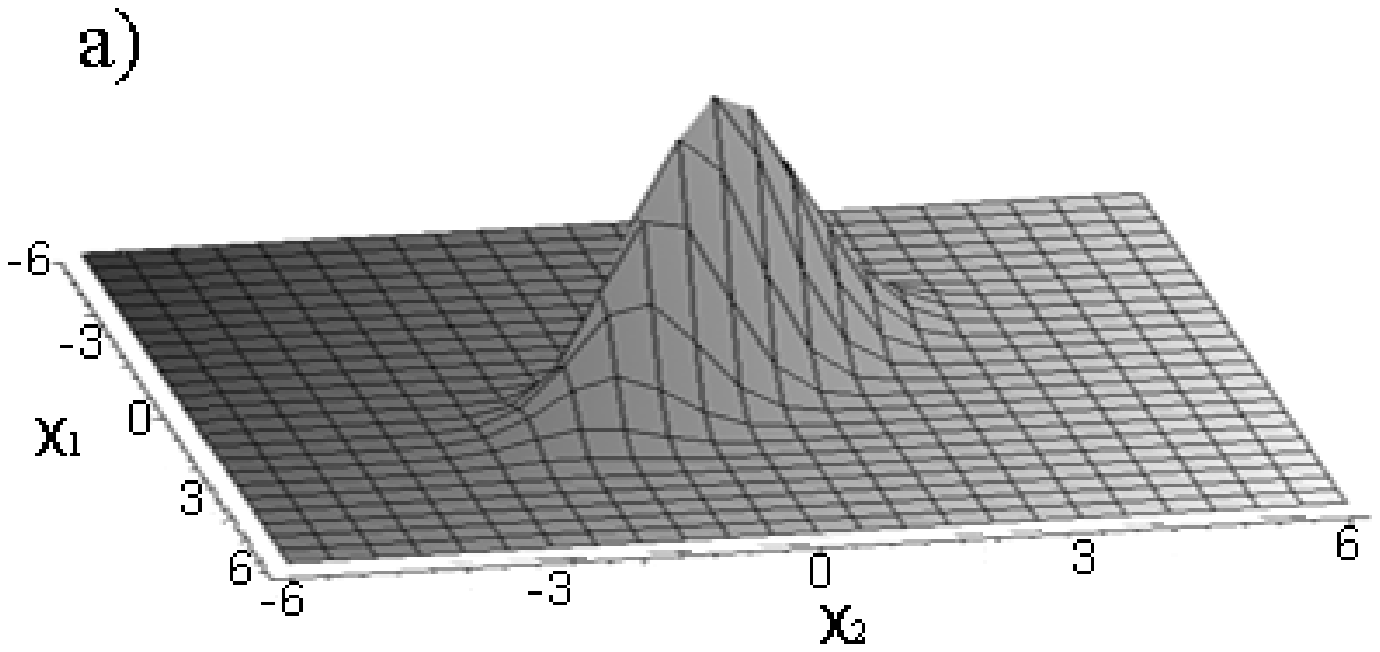}}}
\resizebox*{8.0cm}{5cm}{\rotatebox{0}{\includegraphics{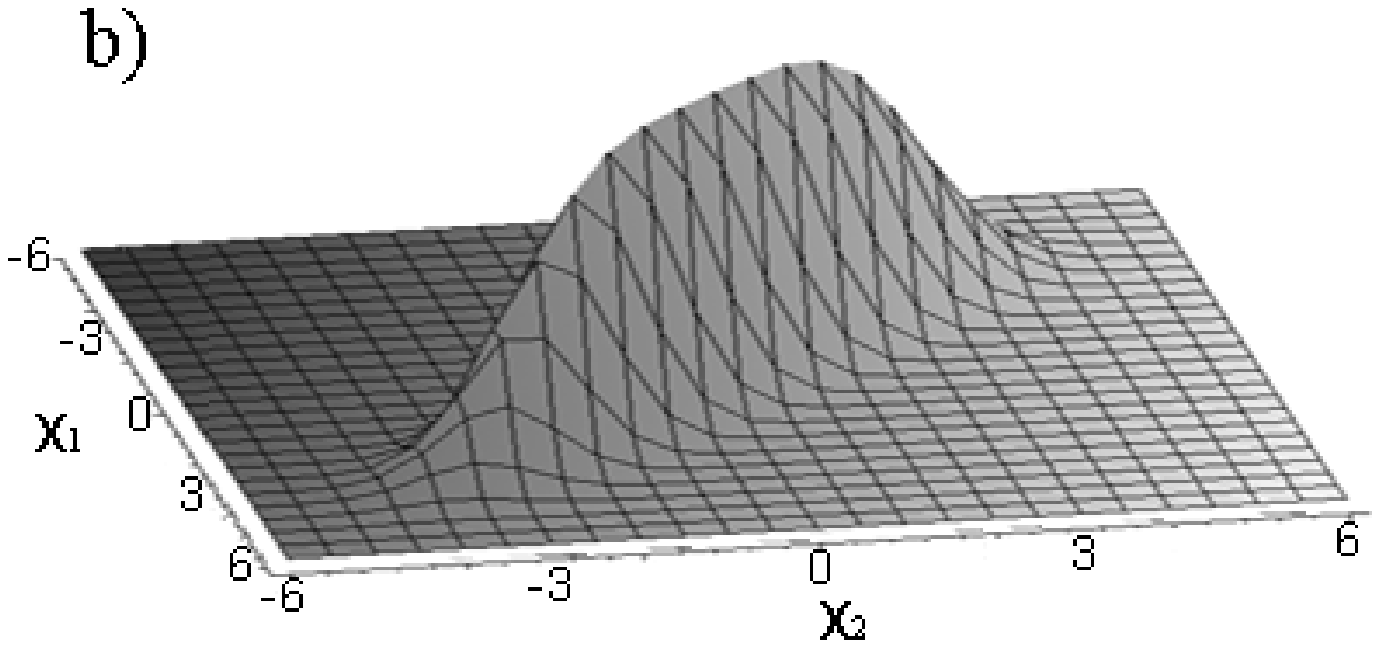}}}
\resizebox*{8.0cm}{5cm}{\rotatebox{0}{\includegraphics{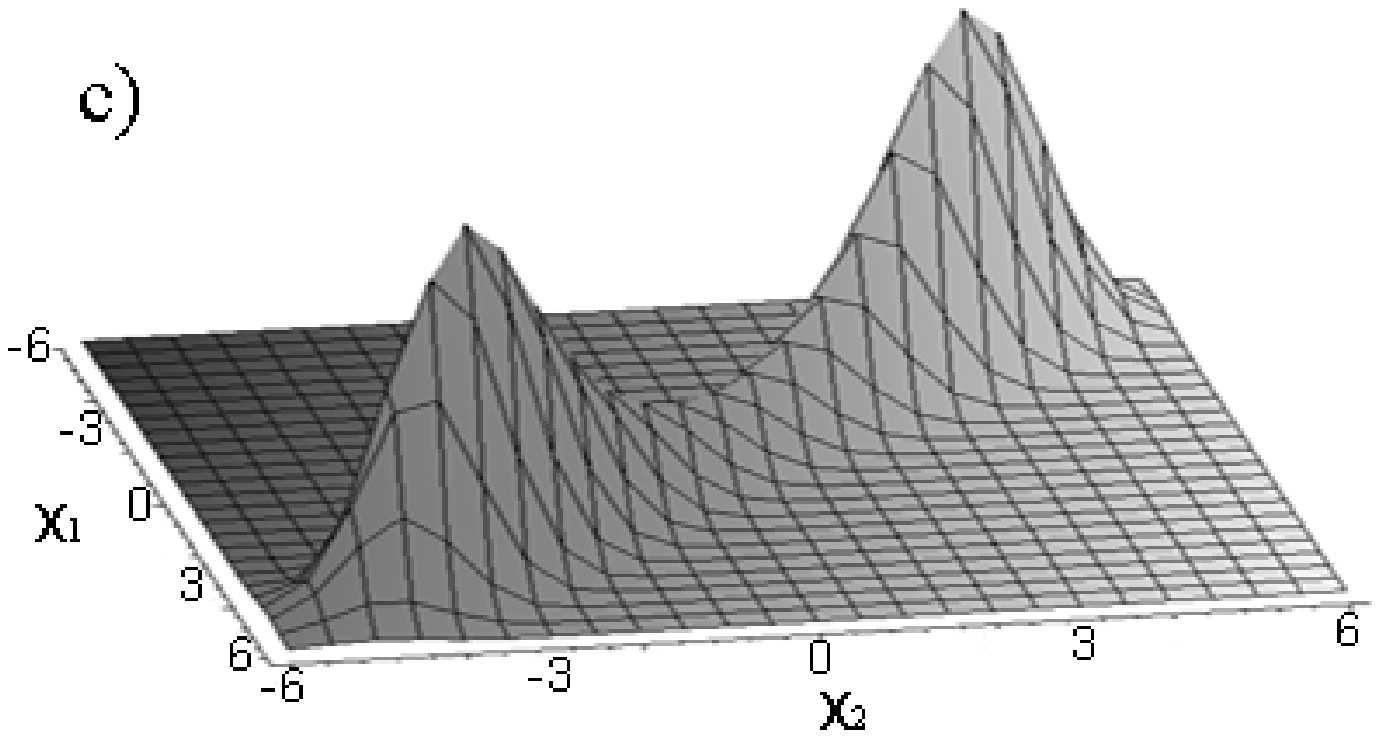}}}
\caption{Conditional Wigner distribution for $y_{1}=0$ and $y_{2}=0$, 
a) below ($\mu = 0.5$); b) at ($\mu = 1.0$); and c) above threshold 
($\mu = 1.5$). In all cases $g^{2}=0.01$.}
\end{figure}

For the nonlinear media employed in most OPO devices, the coupling 
parameter $g\ll 1$. Therefore, the first two terms in the 
exponent of the Wigner distribution governs the OPO behaviour 
below threshold. They are enough to predict the usual 
nonclassical features. In fact, it is evident now that the usual 
linearized approach to quantum noise 
in below-threshold OPOs is equivalent to completely neglect the 
$g^2$ term in the distribution. However, it is important to notice 
that for $\mu\geq 1$, the Wigner distribution becomes divergent if 
we neglect the $g^2$ term. This means that this term plays a 
crucial role for a complete description of the OPO behaviour 
at and above threshold. 

Additional insight is provided by a careful investigation of the marginal 
distribution obtained by integrating $W(x_{1}, y_{1},x_{2}, y_{2})$ with 
respect to one mode variables. Of course, due to the symmetry of $W$, the 
form of the resulting marginal distribution is independent of the mode 
being traced out. The marginal distribution for mode 2 becomes:
\begin{eqnarray}
&&W\left(x_2,y_2\right)=2\pi {\cal{N}}\times
\label{marginal}\\ 
&&\exp\left(\frac{-\left(x_{2}^2+y_{2}^{2}\right)
\left(1+g^2\left(1+x_{2}^2+y_{2}^{2}\right)-\mu^2\right)}
{2 \left( 1+g^2\left(x_{2}^{2}+y_{2}^{2}\right)\right)^2}\right)
\nonumber
\end{eqnarray}
It gives the statistical properties of isolated measurements on mode 2.
The variances given by this marginal distribution are larger than those 
of the vacuum state. This excess noise can be seen as a consequence of 
information loss when one looks only at part of the whole system.
Fig. (2) presents the marginal distribution below, at and above threshold. 
Above threshold the marginal distribution is clearly peaked out of the 
phase space origin, what is a consequence of the macroscopic amplification 
of the down-converted fields. However, the distribution presents radial 
symmetry indicating complete phase uncertainty in each individual mode.

\begin{figure}[ht]
\resizebox*{8.0cm}{5cm}{\rotatebox{0}{\includegraphics{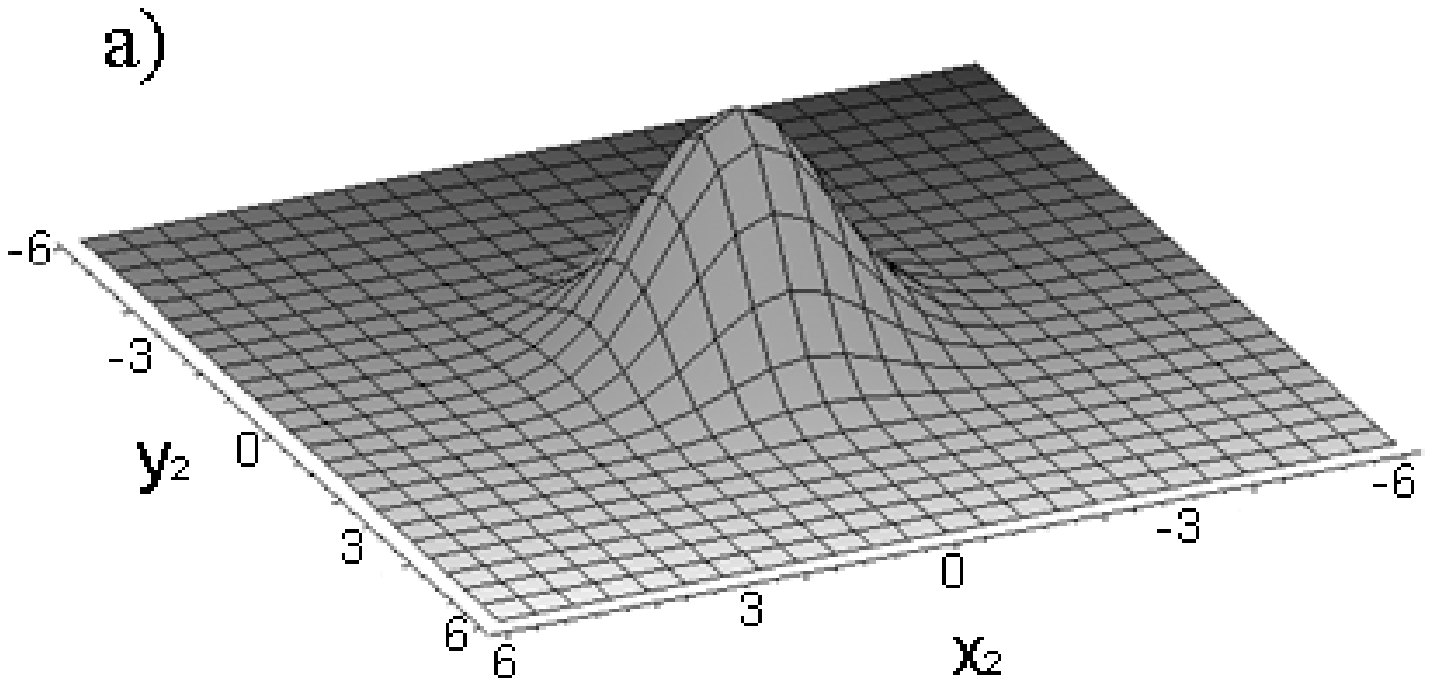}}}
\resizebox*{8.0cm}{5cm}{\rotatebox{0}{\includegraphics{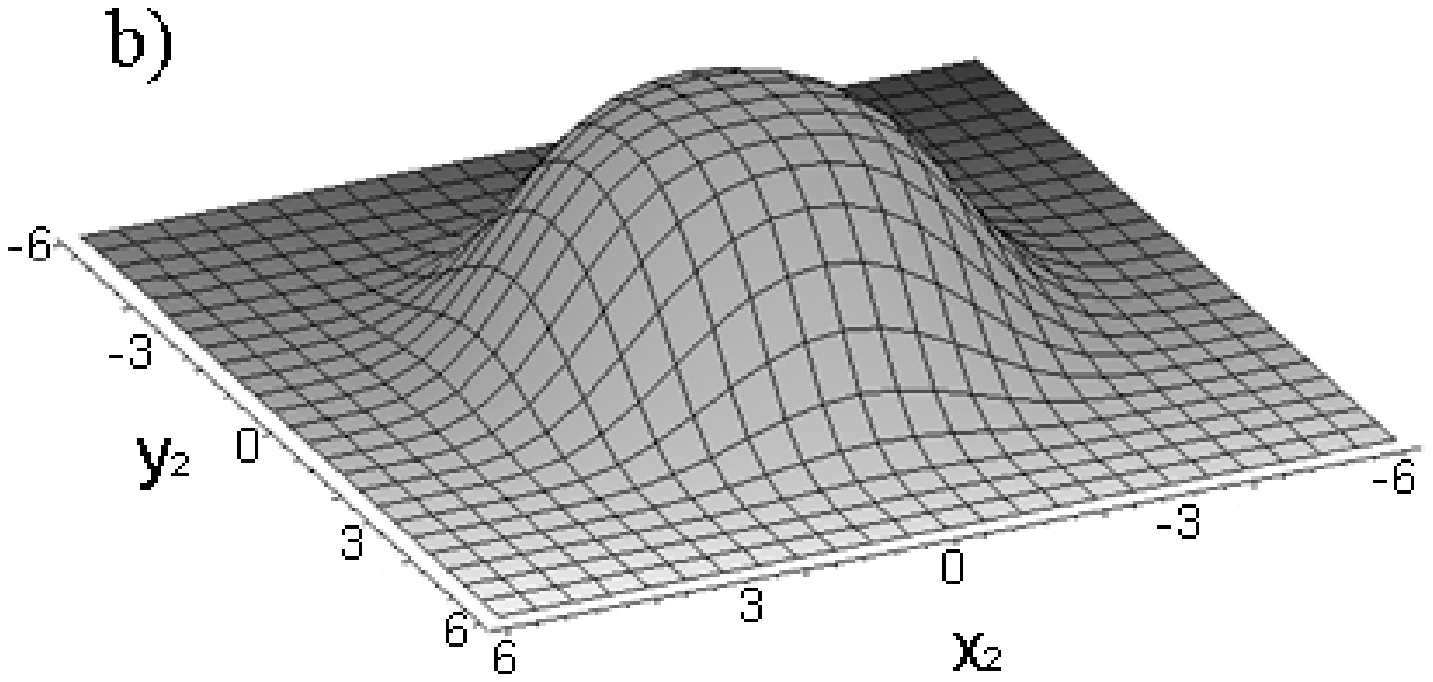}}}
\resizebox*{8.0cm}{5cm}{\rotatebox{0}{\includegraphics{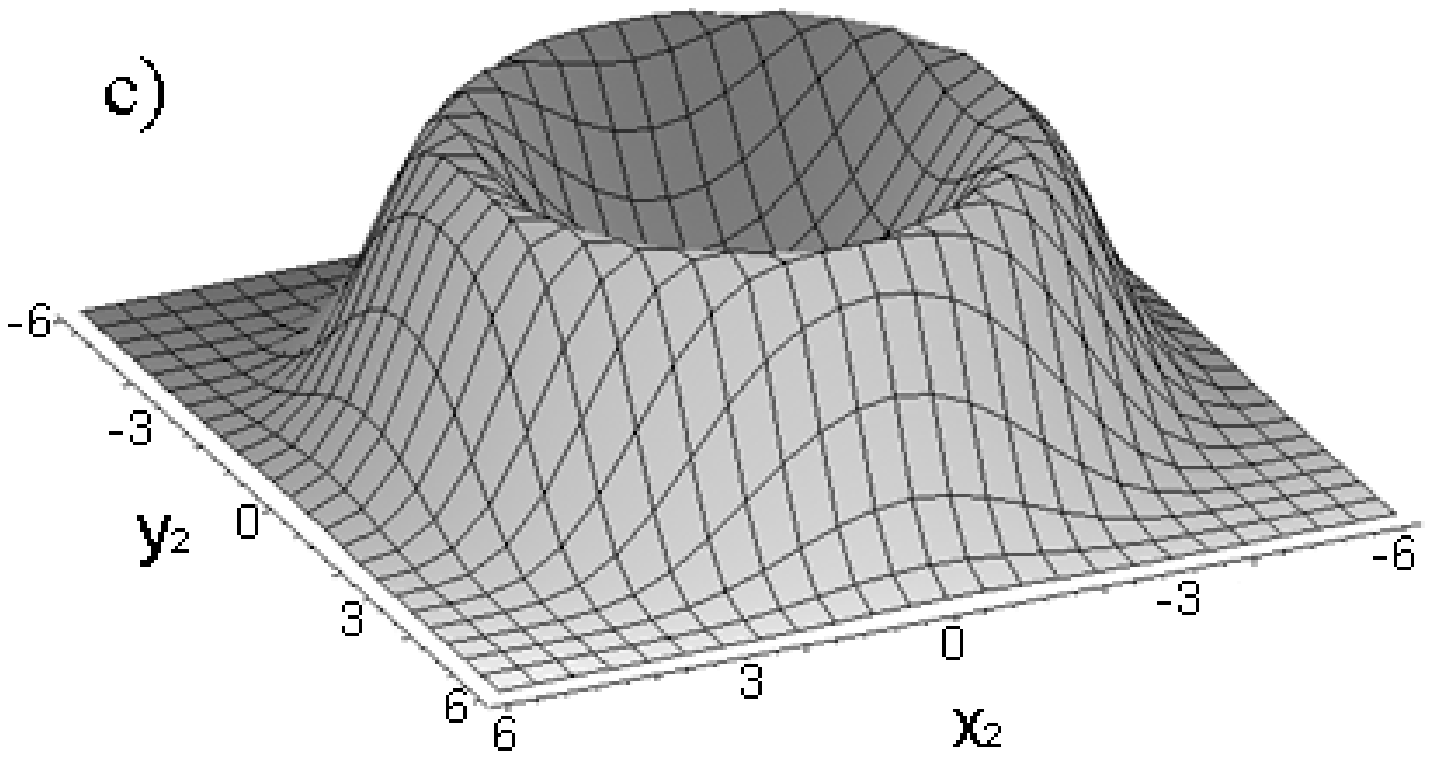}}}
\caption{Marginal distributions: a) below threshold, $\mu = 0.8$; 
b) at threshold, $\mu = 1.0$; 
c) above threshold, $\mu = 1.2$. In all cases $g^{2}=0.01$.}
\end{figure}

\section{Comparison with the linearized quantum approach} 

Almost the totality of theoretical works devoted to the analysis of 
quantum noise in optical systems rely on linearization of the small 
quantum fluctuations around the macroscopic steady state mean values. 
It is worth to notice that this procedure has limited validity, especially 
around the threshold critical point where quantum fluctuations may 
become comparable to the mean values. In order to evidence the 
breakdown of the linearized theory, we shall compare its results with 
those obtained from the solution of the Fokker-Planck equation (\ref{10}). 

The usual approach given to quantum noise in the literature is obtained 
from first order stochastic equations of motion \cite{kaled,villar}.
These equations are often used to predict squeezing in a linearized 
fluctuation analysis. They are non-classical in the sense that they 
can describe states without a positive Glauber-Sudarshan P-distribution, 
but correspond to a Gaussian Wigner distribution. 

We now find it useful to introduce combined field quadratures, as in 
two-mode approaches used previously~\cite{Caves}. These combined quadratures 
are the Einstein-Podolsky-Rosen (EPR) variables used to characterize continuous 
variable entanglement between the down converted fields.  
They are defined as
\begin{equation}
x_{\pm} = \frac{x_{1} \pm x_{2}} {\sqrt{2}}, \;\;\;\;\;\; 
y_{\pm} = \frac{y_{1} \pm y_{2}} {\sqrt{2}}.
\label{eq:newquads}
\end{equation}
These quantities correspond to the squeezed and anti-squeezed combined 
quadratures obtained in the linearized theory. 
>From the first order stochastic equations one can easily obtain the 
steady state variances of the EPR variables \cite{kaled,villar}:
\begin{eqnarray}
\langle x_{+}^{2}\rangle &=& \langle y_{-}^{2}\rangle = \frac{1}{1-\mu}
\nonumber\\
\langle x_{-}^{2}\rangle &=& \langle y_{+}^{2}\rangle = \frac{1}{1+\mu}
\label{eq:variancesbelow}
\end{eqnarray}
for below threshold operation, and
\begin{eqnarray}
\langle x_{+}^{2}\rangle &=& \langle y_{-}^{2}\rangle = \frac{1}{\mu-1}
+ \frac{1}{g^2}(\mu-1)
\nonumber\\
\langle x_{-}^{2}\rangle &=& \langle y_{+}^{2}\rangle = \frac{1}{2}
\label{eq:variancesabove}
\end{eqnarray}
for above threshold operation. 

Now let us briefly discuss the predictions of the linearized approach 
and its validity. The quadratures $x_-$ and $y_+$ exhibit 
the expected squeezing, going from the vacuum fluctuations for zero 
pump until 50\% intracavity squeezing (which corresponds to perfect 
squeezing outside the cavity) at and above the oscillation threshold.
The unsqueezed quadratures $x_+$ and $y_-$ present divergent behaviour 
around threshold, which is certainly not physical. In fact, as we 
shall see next, the steady state solution of the Fokker-Planck equation 
(\ref{10}) gives a well behaved Wigner distribution which does not 
display any divergences. This Wigner distribution will also put limits 
on the amount of squeezing attainable.

In order to investigate the two-mode entanglement directly from the Wigner 
distribution, we now write it in terms of the EPR variables.
Below threshold, where the linearization of the equation of motion is valid, 
we can neglect the $g^{2}$ term, and the distribution can be approximated as 

\begin{eqnarray}
&&W_L(x_{+},y_{+},x_{-},y_{-}) = {\cal{N}} 
exp \left\{-\frac{1}{2}\left[ \left(1+\mu \right) x_{-}^{2} +\right.\right.
\nonumber\\ 
&&\left.\left.\left(1+\mu \right)y_{+}^{2}+ \left(1-\mu \right) x_{+}^{2}+ \left(1-\mu \right) y_{-}^{2} 
\right] \right\}\;. 
\label{weprL}
\end{eqnarray}

With this expression we can easily calculate any moment of the distribution. For instance, we easily 
reobtain the results of Eqs.(\ref{eq:variancesbelow}) for the intracavity noise squeezing in the combined 
quadratures.
As we mentioned before, at threshold we have perfect external squeezing while the unsqueezed combined 
quadratures blow up showing the failure of linearization \cite{kaled}. 
Moreover, it is easy to characterize this state as 
entangled by using the Duan-Simon criterion \cite{simon,duan}. Above threshold we can also calculate the moments 
and we find that this criterion shows the suficient condition to characterize this state as entangled, 
in spite of not being a Gaussian state in that regime.

\begin{figure}
\resizebox*{7.0cm}{5.3cm}{\includegraphics{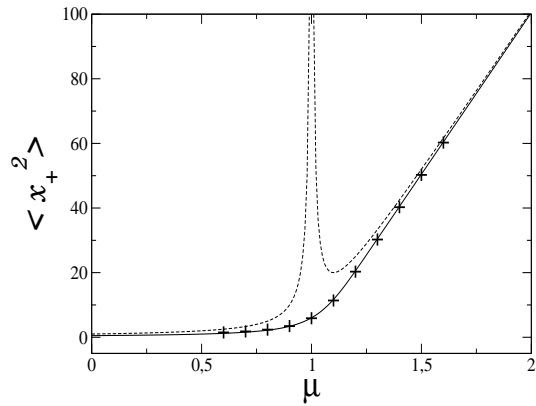}}
\caption{Comparison between the values of the anti-squeezed variable 
$\langle x^2_+\rangle$ obtained from the linearized theory (dashed line) 
with those obtained from the Wigner function of reference \cite{karen} 
(solid line), and the one given by Eq.(\ref{11}) (crosses). In all cases, 
we have set $g^2=0.01$.}
\label{limiar}
\end{figure}

\begin{figure}
\resizebox*{7.0cm}{5.3cm}{\includegraphics{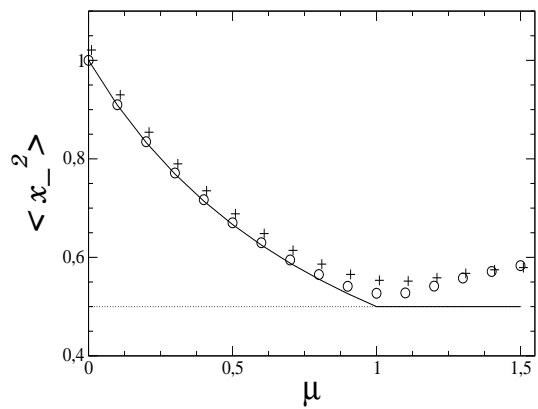}}
\caption{Comparison between the values of the squeezed variable 
$\langle x^2_-\rangle$ obtained from the linearized theory (solid line) with 
those obtained from the Wigner function of Ref. \cite{karen} (circles), 
and the one given by Eq.(\ref{11}) (crosses). In all cases, we have set $g^2=0.01$.}
\label{limiar2}
\end{figure}

It is also interesting to compare the results obtained for $\langle x^2_+\rangle$ 
with the linearized theory (both below and above threshold) with those obtained 
from the Wigner function given by Eq.(\ref{11}). This is done in Fig.(\ref{limiar}). 
While the linearized approach gives a divergent behavior at the oscillation 
threshold, the results obtained from the truncated Wigner function gives a more 
realistic smooth behavior in agreement with the full quantum solution of 
Refs.\cite{karen,mcneil,singh}. 
The results for the squeezed quadrature $\langle x^2_-\rangle$ are shown in Fig.\ref{limiar2} 
where, again, the results given by the truncated Wigner distribution agree with those obtained 
from the full quantum solution. Below threshold, the linearized approach also agrees with the 
full quantum theory and the truncated Wigner approach, but an important deviation can be 
observed above threshold where the linearized theory gives $\langle x^2_-\rangle=0.5$ for 
any pump power. The same results are obtained for the other squeezed variable 
$\langle y^2_+\rangle$, so that the Duan-Simon separability criterion \cite{simon,duan} shows 
that the two-mode quantum state is entangled. However, it is important to notice that both 
the full quantum approach and the truncated Wigner distribution predict a less pronounced 
violation of the criterion above threshold.

In Refs.\cite{mcneil,singh} Fokker-Planck equations were derived 
for the complex-P \cite{mcneil} and positive-P \cite{singh} representations, 
and stationary solutions were obtained for the corresponding distributions. 
Since these Fokker-Planck equations were derived without any truncation, 
we can consider the stationary distributions so obtained as \textit{exact}. 
In Ref.\cite{karen} an exact stationary Wigner function was mapped from 
the complex-P distribution of Ref.\cite{mcneil} and expressed in terms of 
Bessel functions. It is important to remark that the simple Wigner function 
presented here provides a very good approximation for those exact solutions, 
as can be seen in Figs.\ref{limiar} and \ref{limiar2}, even for a rather large value for the 
nonlinear coupling $g^2$. 

\section{Conclusion}

By truncating the evolution equation for the two-mode Wigner function of 
an optical parametric oscillator, we derived a Fokker-Planck equation which 
admits a simple potential solution in any operation regime of the OPO. 
For the unsqueezed quadratures, this solution 
does not present the unphysical divergent behavior of the linearized theory on the 
oscillation threshold. Moreover, it provides good agreement with exact quasiprobabilities 
already available in the literature, both for the squeezed and unsqueezed quadratures. 
While these exact solutions are given as infinite series, the potential solution 
of the truncated Wigner distribution presents a simple form allowing for an 
easier visualization of the phase space distribution.

\begin{acknowledgments}

This work was supported by the Instituto Nacional de Ci\^encia e Tecnologia de 
Informa\c c\~ao Qu\^antica (INCTIQ - CNPq - Brazil), Coordena\c c\~ao de 
Aperfei\c coamento de Pessoal de N\'{\i}vel Superior (CAPES - Brazil) and Funda\c c\~ao 
de Amparo \`a Pesquisa do Estado do Rio de Janeiro (FAPERJ).

\end{acknowledgments}

%



\begin{thebibliography}{xxxx}
%
\bibitem{kinsler} P. Kinsler and P. D. Drummond, Phys. Rev. A {\bf 43}, 6194 (1991).
%
\bibitem{kaled2} P. D. Drummond and K. Dechoum, Phys. Rev. Lett. {\bf 95}, 083601 (2005).
%
\bibitem{karen} K. V. Kheruntsyan, and K. G. Petrosyan, Phys. Rev. A {\bf 62}, 015801 (2000).
%
\bibitem{simon} R. Simon, Phys. Rev. Lett. {\bf 84}, 2726 (2000).
%
\bibitem{duan} L. M. Duan, G. Giedke, J. I. Cirac, and P. Zoller, 
Phys. Rev. Lett. {\bf 84}, 2722 (2000).
%
\bibitem{mcneil} K. J. McNeil and C. W. Gardiner, 
Phys. Rev. A {\bf}, 28, 1560 (1983).
%
\bibitem{carmichael} H. J. Carmichael, \textit{Statistical Methods in Quantum 
Optics 1} (Springer, Berlin, 1999). 
%
\bibitem{nos} B. Coutinho dos Santos, K. Dechoum, A. Z. Khoury, L.F. da Silva, and M.K. Olsen 
\pra \textbf{72}, 033820, (2005).
%
\bibitem{ndturco}K. Dechoum, P. D. Drummond, S. Chaturvedi, and M. D. Reid, 
\pra, \textbf{70}, 053807 (2004).
%
\bibitem{Caves} C. M. Caves and B. L. Schumaker, Phys. Rev. A \textbf{31}, 
3068 (1985); B. L. Schumaker and C. M. Caves, \textit{ibid.} \textbf{31}, 3093 
(1985).
%
\bibitem{Kimble} H. J. Kimble, \textit{Fundamental systems in quantum 
optics} edited by J. Dalibard, J. M. Raimond and J. Zinn-Justin (1992).
%
\bibitem{villar} A. S. Villar, K. N. Cassemiro, K. Dechoum, A. Z. Khoury, 
M. Martinelli, P. Nussenzveig, J. Opt. Soc. Am. B {\bf 24}, 249 (2007).
%
\bibitem{kaled} K. Dechoum, P. D. Drummond, S. Chaturvedi and M. D. Reid, 
Phys. Rev. A {\bf 70}, 053807 (2004).
%
\bibitem{gardiner} C. W. Gardiner, \textit{Handbook of Stochastic Methods} 
(Springer, Berlin, 1985).
%
\bibitem{singh} R. Vyas, S. Singh, Phys. Rev. Lett. {\bf 74}, 2208 (1995).
%
\end{thebibliography}
\end{document}